\documentclass[manuscript,screen,nonacm]{acmart}
\usepackage{graphicx}
\usepackage{lipsum}
\usepackage{listings}
\usepackage{subcaption}
\usepackage{todonotes}
\usepackage{verbatim}
\graphicspath{{./Images/}}
\usepackage{fancyhdr}

\renewcommand\footnotetextcopyrightpermission[1]{}
\setcopyright{none}

\usepackage{xcolor}

\begin{document}

\title{Do Cloud Developers Prefer CLIs or Web Consoles?}
\subtitle{CLIs Mostly, Though It Varies by Task}

\author{Cora Coleman}
\email{ccoleman@eng.ucsd.edu}
\affiliation{
  \institution{University of California, San Diego}
  \streetaddress{9500 Gilman Drive}
  \city{La Jolla}
  \state{California}
  \country{USA}
  \postcode{92093}
}

\author{William G. Griswold}
\email{wgg@eng.ucsd.edu}
\affiliation{
  \institution{University of California, San Diego}
  \streetaddress{9500 Gilman Drive}
  \city{La Jolla}
  \state{California}
  \country{USA}
  \postcode{92093}
}

\author{Nick Mitchell}
\email{nickm@us.ibm.com}
\affiliation{
  \institution{IBM T.J. Watson Research Center}
  \streetaddress{1101 Kitchawan Road}
  \city{Yorktown Heights}
  \state{New York}
  \country{USA}
  \postcode{10598}
}

\begin{abstract}
Despite the increased importance of Cloud tooling, and many large-scale studies of Cloud users, research has yet to answer what tool modalities (e.g. CLI or web console) developers prefer. In formulating our studies, we quickly found that preference varies heavily based on the programming task at hand. To address this gap, we conducted a two-part research study that quantifies \textit{modality preference as a function of programming task}. Part one surveys how preference for three tool modalities (CLI, IDE, web console) varies across three classes of task (CRUD, debugging, monitoring). The survey shows, among 60 respondents, developers most prefer the CLI modality, especially for CRUD tasks. Monitoring tasks are the exception for which developers prefer the web console. Part two observes how four participants complete a task using the kubectl CLI and the OpenShift web console. All four participants prefer using the CLI to accomplish the task. \end{abstract}


\maketitle

\section{Introduction}

Developing for the Cloud can be arduous\cite{avram2014, dillon2010}. The set of APIs is large and constantly changing. The Cloud is inherently asynchronous and distributed. Tools can help, but are often presented as rigidly separated \emph{modalities}. These tools employ either low-level and flexible conversational interfaces, or high-level and limited direct-manipulation interfaces \cite{li2020, hutchins1985}. This rigid separation can harm productivity, requiring one either to cope with low-level tools, dance around the limitations of the high-level tools, or pay the cognitive burden of frequent context switches between the two \cite{abad2018, myers2000}.

This paper aims to foster constructive dialog around better tools and to understand the important reasons and contextual factors that motivate developer tool preference \cite{cangiano2009}. Cloud tools run the gamut of modalities. Every major Cloud provider offers command line interfaces (CLIs), Integrated Development Environment (IDE) plugins, and browser-based (web) consoles \cite{avram2014}. These tools belong to categories of modalities that drive different user interactions. Industry developers are on-boarded to use a certain set of Cloud technologies and tools, however research has yet to answer what modalities of tools developers prefer to use.

The goal of all of these tools is help to developers interface with a set of Cloud APIs. This paper focuses on one popular family of APIs known as Kubernetes. The Kubernetes framework is an open-source system for automating the deployment, scaling, and management of containerized applications \cite{k8s2021}. When using Kubernetes, Cloud developers must choose a development tool to support their “inner loop”, i.e. coding, deployment of prototypes, and debugging any problems that arise. Cloud-based environments support these development tasks and also offer access to APIs, microservices, and DevOps \cite{ibm2021}. Commonly used Kubernetes development tools include the \texttt{kubectl} CLI \cite{kubectl2022}, the VSCode IDE \cite{vscode2022}, and a variety of web consoles (such as the Kubernetes Dashboard or the OpenShift Console) \cite{dashboard2022, openshift2022}.

There are few research studies on developer tools and tools for novices that examine tool preference in the context of the types of programming tasks that developers commonly iterate on or within the user population of Cloud developers \cite{abad2018, wang2017}. This paper contributes to filling this research gap with two user studies. First, it presents survey findings from 60 Cloud developers. The survey allows us to pivot our analysis along three categorical dimensions:
\begin{enumerate}
\item \textbf{Task} being performed: CRUD, debugging, monitoring.
\item \textbf{Modality} of tool used: CLI, IDE, web console.
\item \textbf{Expertise} in Kubernetes: beginner through advanced.
\end{enumerate}

In discussions with industrial designers, the implicit (and largely informal) assumption seems to be that beginners will always prefer web consoles, and experts will prefer CLIs \cite{chen2007}. In the studies presented in this paper, we have found that this is not the case, that modality preference is not primarily a function of one's level of expertise. Rather, this paper shows that a primary factor influencing preference is the task being performed \cite{abad2018, wang2017}. Furthermore, the frequency and types of tasks of Cloud developers is not answered in prior research. We have found that multiple types of tasks are performed each day, supporting prior research that developers are often interrupted and start one task and move on to another before finishing the first, known as multitasking or interruption \cite{abad2018, gonzalez2004, vasilescu2016}.

Our survey of Cloud developers shows that, regardless of their level of experience, developers prefer the CLI. The exception for which developers prefer a different tool modality was found to be a web console for monitoring tasks. To dig deeper and uncover the reasons that motivate developer preferences, we observed developers using different tool modalities to complete a programming task. The second contribution of the paper presents findings from a comparative user study of four participants that each completed a task using the kubectl CLI and the OpenShift web console. The observational study corroborates the findings from the survey in that all four participants preferred to use the CLI when performing a task.

\section{An Example Cloud Developer Scenario}

\begin{figure*}
    \begin{subfigure}[b]{0.45\textwidth}
      \centering
      \includegraphics[width=\linewidth]{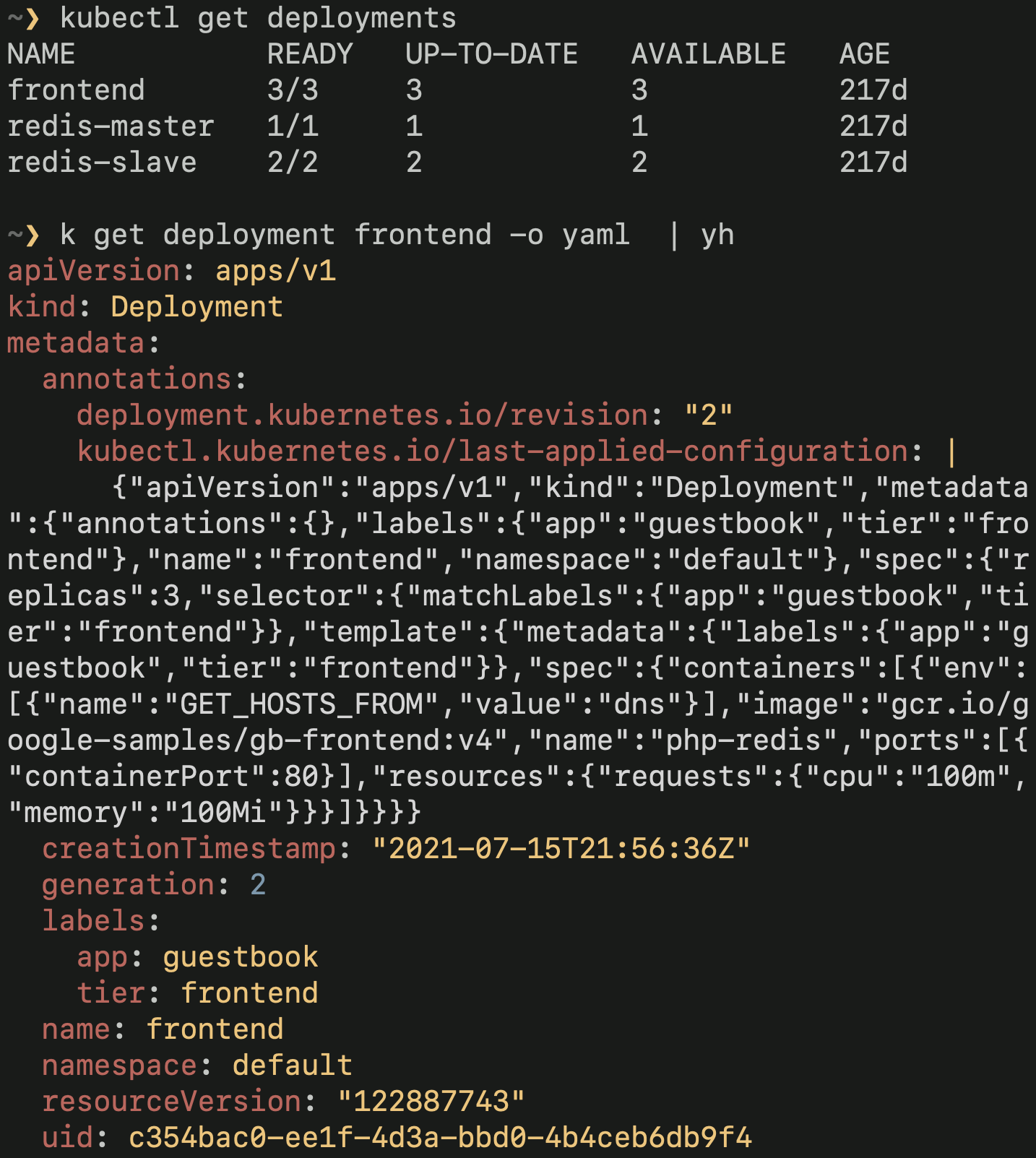}
      \caption{The \texttt{kubectl} CLI uses a conversational interface. It allows copy/paste of commands. Switching between tasks can leverage custom scripts and command history navigation.}
      \label{fig:kubectl}
    \end{subfigure}
    \quad
    \begin{subfigure}[b]{.51\textwidth}
      \centering
      \includegraphics[width=\linewidth]{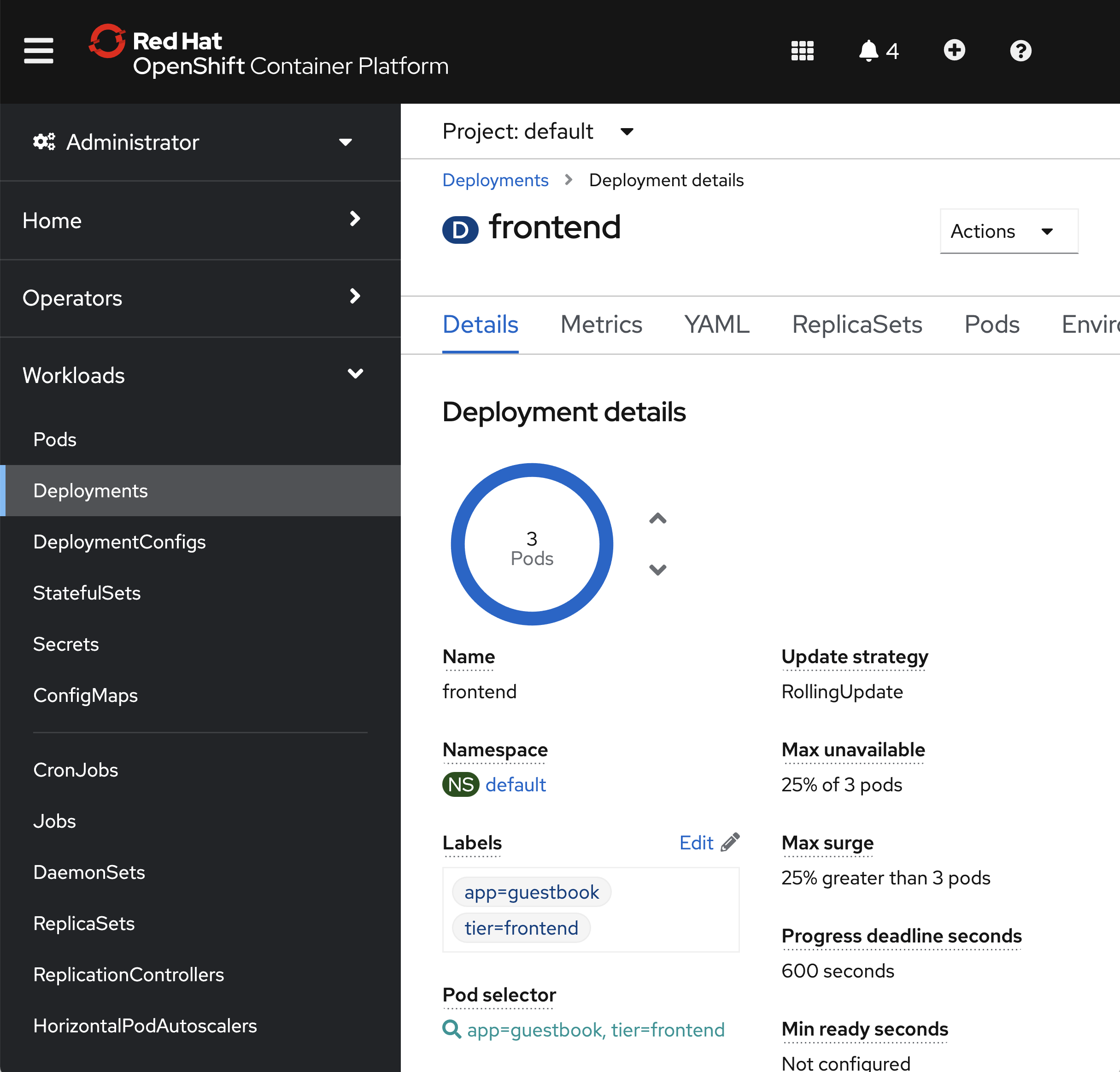}
      \caption{The OpenShift console uses a direct-manipulation interface. It requires clicks to perform navigation, and is presented as a series of modal pages. This view shows the details page for a Deployment resource. Switching between views often requires multiple clicks.}
      \label{fig:openshift}
    \end{subfigure}
    
  \caption{Using two Cloud tools to list and view the details of Deployment resources in a Kubernetes cluster.}
  \label{fig:}
\end{figure*}

A Cloud developer's day typically involves multiplexing their time between several categories of tasks: CRUD\footnote{CRUD: Create Read Update Delete, i.e. tasks that inspect or modify an application resource.}, debugging, and monitoring. Cloud applications are complex enough, with multi-process distribution and scale to the point that blurs the classical lines between development and operations. Developing with this complexity can no longer be done in silos~\cite{netflix}. Thus, it is important to capture this cross-task nature when evaluating tool preference.

To further illustrate the challenges that exist in Cloud development, consider the following scenario that involves a simple monitoring task. This scenario is one based on the results presented later in this paper, and typical of a daily driving experience in Cloud development:

Star is an experienced Java developer with less than one month of experience developing for the Cloud and considers herself a beginner. Star learned the basics by referencing tutorials in blog posts, and by browsing platform documentation such as is available on the Kubernetes web site (kubernetes.io). This documentation is focused on teaching by rote: it gives examples of CLI commands that users can copy and paste into their terminal in order to accomplish common tasks. In this way, she learned how to start a development cluster on her laptop, and then how to deploy and manage applications through use of the \texttt{kubectl} CLI. See~\autoref{fig:kubectl} for an example of this CLI experience.

Many times each day, Star needs to update her deployed application with her latest changes (a CRUD task). She may need to inspect the logs, if the update causes the application to misbehave (a debugging task). Threaded through each of these tasks, she will routinely be checking the status of her deployed resources. For example, after pushing an update, she may need to wait for the resources to be "green", meaning that the deployment is complete, and the distributed set of application resources are responsive to pings. Or, if there is a problem, she may need to identify which of the resources are "red". These are monitoring tasks.

\begin{figure}
    \centering
    \includegraphics[width=0.4\linewidth]{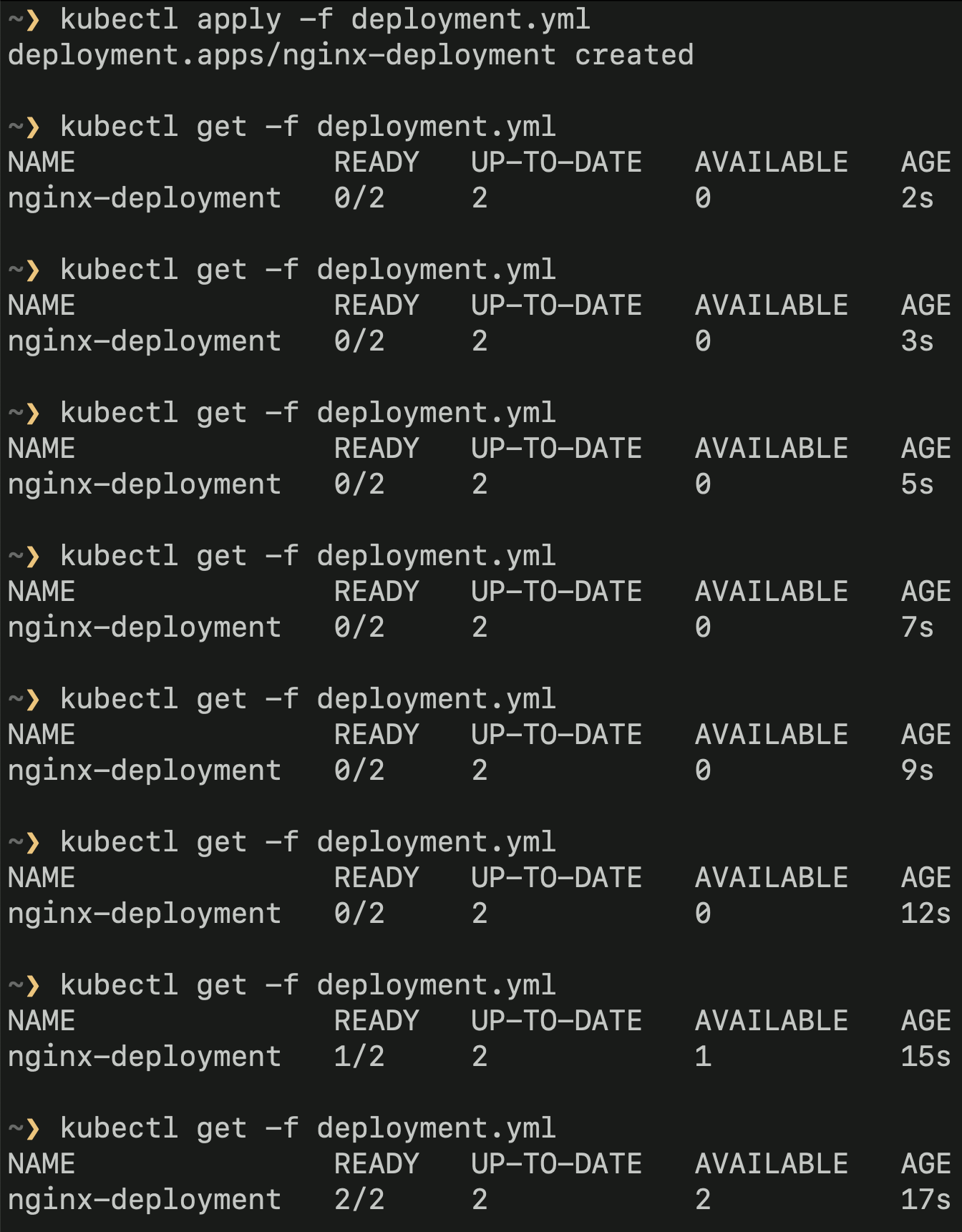}
    \caption{Our participant "Star" uses the \texttt{kubectl} CLI. First she updates her application via \texttt{kubectl apply}, then polls via \texttt{kubectl get} until it is ready --- as indicated by the "READY 2/2" cell. Note from the AGE column how she waited 17 seconds.}
    \label{fig:kubectl-spam}
\end{figure}

Let us focus on a monitoring task. Star has just deployed an update, and needs to wait for them to be ready. Using the CLI, Star issues a command to query the status of her Deployment resource; this Kubernetes resource type roughly corresponds to an application. She repeats this command until she sees that the resources are ready.~\footnote{You may be surprised that Star is manually repeating the command, rather than using the watch capability of the kubectl CLI. None of our comparative user study participants employed this feature.}
\autoref{fig:kubectl-spam} shows a screenshot of such a session.

Star notices that another developer on her team uses the OpenShift (OS) web console for monitoring application progress and that it provides a clear visualization throughout the application creation process; see~\autoref{fig:openshift}. Star decides to use both the kubectl CLI and the OS web console to create and monitor her applications, respectively. Star finds that while OS is great for monitoring, it can be cumbersome to switch contexts between the CLI and web console. For example, the user interactions are completely different in each interface. Star must switch between the keyboard and the mouse instead of sticking to one input method to navigate through the CLI commands and console menus.

The main takeaways from this example are that conversational and direct-manipulation interfaces found in CLIs and web consoles, respectively, help developers to accomplish different types of tasks. Additionally, developers often work on multiple types of tasks each day and look to optimize their performance by using tools with modalities that suit the task at hand. This can lead to cognitive burden from context switching between different programming tasks and development environments. Seems like a perfect candidate example for graphics, however we have found that this is not always the case.

\section{Related Work}

There have been few studies focused on understanding Cloud developer tooling perspectives in the context of the types of programming tasks developers complete. The most relevant work in understanding how Cloud developers use tools evaluates how developers complete common programming task types in an IDE \cite{wang2017}. Developer surveys such as the CNCF~\footnote{CNCF: Cloud Native Computing Foundation, an open-source consortium that manages many projects, including Kubernetes. https://www.cncf.io/} report on the increased usage of Cloud tooling and don’t clarify which tools developers prefer to use or the types of programming tasks that developers are working on \cite{cncf2021}. Other developer surveys from open source special interest groups (SIGs) don't ask about the programming task context in which the tool is used aside from what tools are used for CI/CD \cite{sigapps2018}.

\subsection{Systems Administrators and Monitoring}

Research in understanding developer tooling outside of the Cloud has focused on the user population of system administrators, specifically for monitoring tasks such as intrusion detection \cite{thompson2007, barrett2004, voronkov2019}. Research on graphical and textual tool modalities for intrusion detection have revealed some of the relative strengths and weaknesses of each and led to recommendations for a hybrid tool \cite{thompson2007}. The textual tool was found to allow users to better analyze data through flexible and efficient commands while the graphical tool was found to better support the discovery of new attacks by providing a system overview of the current state of the network. 

A field study of systems administrators revealed how automation and scripting are crucial to their work \cite{barrett2004}. While CLIs offer more support on this end than graphical user interfaces (GUIs), they do so with a higher learning curve, less ease of use, and provide less insight to the system interactions. A survey of systems administrators identified that most prefer GUIs over CLIs for managing firewalls \cite{voronkov2019}. The CLI’s most reported strength was its flexibility and efficiency and the GUIs most reported strength was its usability.

\subsection{Tool Modalities for Learning}

Investigating software tools and specifically, CLIs and their accessibility have revealed some reasons for inefficiency that are universally experienced, though with a steeper learning curve for novice programmers \cite{gandhi2020lightening, myers2000, sampath2021}. CLIs were found to challenge users to remember strings of commands without context, issue terse commands with limited readability for future comprehension, and reduce their editing space to a single statement prone to error \cite{gandhi2020lightening}. The unstructured nature of text within a CLI also makes them inefficient to use with screen readers \cite{sampath2021}. While a lesser learning curve has been observed with GUIs, transitioning to learning using a CLI resulted in more frustration and less success in completing programming tasks than transitioning from learning using a CLI to a GUI \cite{dillon2012comparing, unal2021effects}. 

\subsection{Hybrid Tool Modalities}

Hybrid tools have been found to help with learning how to program by reducing cognitive load and leading to increased programming success \cite{unal2021effects}. CLIs were improved to provide customized GUIs that combine the flexibility and efficiency of commands with the usability and discoverability of graphics \cite{vaithilingam2019}. The customized GUI benefited novice programmers by only exposing the most relevant subset of options required for their specific goal \cite{vaithilingam2019}. Similarly, transforming any web page into an example-centric IDE eliminated the complexities that impede novice programmers in desktop-based environments such as software installation and configuration, data file management, data parsing, and Unix-like command-line interfaces \cite{zhang2017}. 

CLIs were improved to provide customized GUIs that combine the flexibility and efficiency of commands with the usability and discoverability of graphics \cite{vaithilingam2019}. The customized GUI benefited novice programmers by only exposing the most relevant subset of options required for their specific goal. Similarly, transforming any web page into an example-centric IDE eliminated the complexities that impede novice programmers in desktop-based environments such as software installation and configuration, data file management, data parsing, and Unix-like command-line interfaces \cite{zhang2017}.

\section{Methodology}

This paper evaluates the following research questions in a two-part study that includes a survey of 60 Cloud developers and comparative user study with four Cloud research developers:
\begin{enumerate}
\item How do different levels of Kubernetes \textbf{experience} (beginner, intermediate, and expert) impact developer tool preferences?
\item How does the \textbf{type of programming task at hand} (CRUD, debugging, and monitoring) impact developer tool preferences?
\item How does the \textbf{tool modality} (textual, graphical, and hybrid) impact developer tool preferences?
\end{enumerate}

A Google Forms survey was used to collect 60 responses from within the IBM Hybrid Cloud research group and Kubernetes open source special interest groups (SIGs). Participants indicated their level of experience with Kubernetes, self-selecting from either beginner, intermediate, or expert. We also asked participants to indicate how frequently they use Kubernetes, selecting from either many times each day, once a day, weekly, monthly or even more rarely, or never. We then conducted a remotely moderated, within-subjects user study of four participants with an average of 20 years of programming experience between them. All four user study participants belonged to the IBM Hybrid Cloud research group and regularly used Kubernetes in their work.

Survey participants were recruited from within the IBM Hybrid Cloud research group and from Kubernetes Special Interest Groups (SIGs) known as sig-cli and sig-usability. Furthermore, the survey link was publicly included in a post on LinkedIn made by a sig-usability leader and on the GitHub readme file of an open source Kubernetes project, Kui. The survey asked Kubernetes developers questions about their level of experience (beginner, intermediate, expert), frequency of working on different programming task types (CRUD, debugging, monitoring), and preferred tool modality for different programming task types (CLI, Web console, IDE).

The results of the developer survey were used to inform the design of the second part of our study, a comparative user study to observe differences in how participants use the kubectl CLI and the OpenShift console to complete the same programming task. We employed a classic ``Think Aloud'' user evaluation method in which participants were asked to think aloud while completing the task and given a prompt to continue if they fell silent for more than 15 seconds \cite{mcdonald2020impact, ericsson1984protocol, fox2011procedures}. Video and audio from the sessions were recorded using WebEx for later transcription and analysis.

Participants filled out a background survey to provide information about their programming experience, use of Kubernetes, and familiarity with the two tools. Participants downloaded the kubectl CLI and the OpenShift Console onto their machines and were given access to an OpenShift cluster to run Kubernetes. After completing the observational study, participants filled out a follow up survey and ranked the tools based on their preference. In the follow up survey, participants were also asked to rate the tools based on their perceived cognitive load using the NASA-TLX scale \cite{hart1988development}. To give us a better understanding of preference, participants also rated the quality of each tool and were asked to give reasons for why they either liked or did not like using each tool. Participants were also asked to share their recommendations for how each tool could be improved.

During the study sessions, participants completed a task using the kubectl CLI and the OpenShift console. The goal of the task was to create an application in a Kubernetes namespace~\footnote{A Kubernetes namespace can be used to isolate applications from each other.}, find the deployment pods that have the string ``magic key'' in their logs, delete the application, and then delete the namespace. Participants were given a GitHub repository with the application files and task instructions. No time limit was set to complete the task. If participants got stuck while completing the task, hints specific to each tool interface were provided in hyperlinks below the task instructions. The participants were randomly assigned such that half of them used the kubectl CLI first and the other half used the OpenShift console first.

\section{Results}
\subsection{Part One: Developer Survey}

A majority of survey respondents (53\%) said that they have an expert level of experience with Kubernetes and even more (80\%) said that they use Kubernetes many times each day. In total, 32 respondents said that they have an expert level of experience, 23 respondents said that they have an intermediate level of experience and five respondents (8\%) said that they have a beginner level of experience. 

\subsubsection{Frequency of Tasks}

\begin{figure*}
    \begin{subfigure}[t]{0.32\linewidth}
        \centering
        \includegraphics[width=\linewidth]{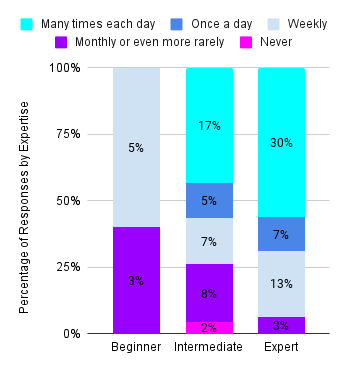}
        \caption{CRUD tasks are most frequently completed on a daily basis by intermediates and experts. No beginners indicated that they complete CRUD tasks more frequently than on a weekly basis.}
        \label{fig:crudxf}
    \end{subfigure}
    \,
    \begin{subfigure}[t]{0.32\linewidth}
        \centering 
        \includegraphics[width=\linewidth]{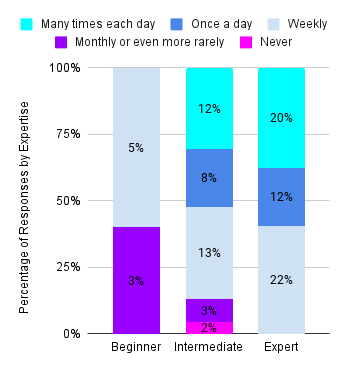}
        \caption{Debugging tasks are most frequently completed on a weekly basis by beginners, intermediates, and experts.}
        \label{fig:debugxf}
    \end{subfigure}
    \,
    \begin{subfigure}[t]{0.32\linewidth}
        \centering
        \includegraphics[width=\linewidth]{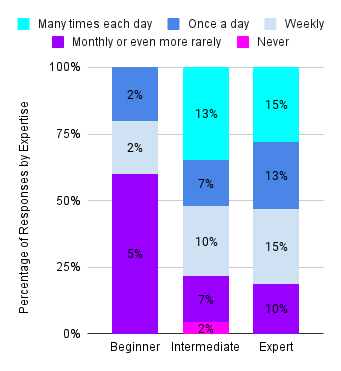}
        \caption{Monitoring tasks are most frequently completed on a daily basis by beginners, intermediates, and experts. The same amount of experts also said that they complete monitoring tasks on a weekly basis.}
        \label{fig:monitorxf}
    \end{subfigure}
    
    \caption{Task frequency from survey participants. The numbers inside each bar represent the fraction of all participants. The height of each bar represents the fraction of that class of users.}
    \label{fig:task-frequency}
\end{figure*}

28 survey respondents (47\%) said that they work on CRUD tasks many times each day, 19 (32\%) said that they work on debugging tasks many times each day, and 17 (28\%) said that they work on monitoring tasks many times each day. 15 respondents (25\%) said that they work on CRUD tasks weekly, 24 (40\%) said that they work on debugging tasks weekly, and 16 (27\%) said that they work on monitoring tasks weekly. One respondent said that they never worked on any of these programming task types, so the task types could have been better explained or we could have included more task types. See ~\autoref{fig:task-frequency}.

\subsubsection{Tool Modality Preference Across Tasks}

\begin{figure*}
    \begin{subfigure}[t]{0.32\linewidth}
        \centering  
        \includegraphics[width=\linewidth]{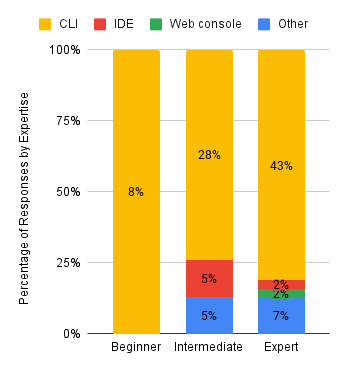}
        \caption{CLIs are preferred by the majority of developers for CRUD tasks, regardless of experience.}
        \label{fig:crudxe}
    \end{subfigure}
    \,
    \begin{subfigure}[t]{0.32\linewidth}
        \centering  
        \includegraphics[width=\linewidth]{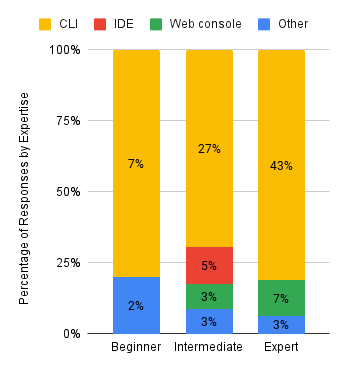}
        \caption{CLIs are preferred by the majority of developers for debugging tasks, regardless of experience.}
        \label{fig:debuggingxe}
    \end{subfigure}
    \,
    \begin{subfigure}[t]{0.32\linewidth}
        \centering
        \includegraphics[width=\linewidth]{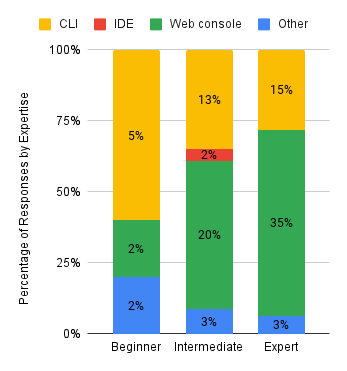}
        \caption{Web consoles are preferred by the majority of intermediate and expert developers, while CLIs are preferred by a majority of beginners for monitoring tasks.}
        \label{fig:monitoringxe}
    \end{subfigure}

    \caption{Modality preference from survey respondents. The numbers inside each bar represent the fraction of all participants. The height of each bar represents the fraction of that class of users.}
    \label{fig:experience-preference}
\end{figure*}

We asked survey participants what kind of tool (CLI, IDE, web console, or other) they prefer to use when working on CRUD, debugging, and monitoring tasks. While a majority of participants said that they prefer to use a CLI when working on CRUD (80\%) and debugging (77\%) tasks, a majority of participants said that they preferred to use a web console to work on monitoring tasks (57\%). See ~\autoref{fig:experience-preference}. 

Overall, we found that the most preferred tool for Kubernetes beginners performing CRUD, debugging, and monitoring tasks is a CLI. Intermediate and expert Kubernetes developers prefer to use a CLI for CRUD and debugging tasks, and a web console for monitoring tasks. One survey respondent said that they preferred to use a combination of a CLI and web console for CRUD and the same for monitoring tasks. Two of the respondents said that they preferred this combination for debugging tasks as well. Lists of the other tools mentioned in survey responses and the tasks for which they are most frequently used can be seen in~\autoref{table:other-combined}.

\begin{table}
    \caption{When survey respondents answered "Other" for their tool preference, the selection of tools they specified also varied by task.}
\begin{subtable}[t]{0.3\linewidth}
\caption{"Other tool" for CRUD tasks.}
\label{table:otherCRUD}
\begin{tabular}{p{0.825\linewidth}c} 
\toprule
Participant response & Votes \\
\midrule
CLI and others (e.g. IDE, depends on specific case) & 1 \\
CLI and web console & 1 \\
CLI, scripting, and automation & 1 \\
Console for browsing multiple objects, CLI for applies and deletes, or piping into other tools & 1 \\
flux and GitHub & 1 \\
Helm charts & 1 \\
infra & 1 \\
kubectl, Grafana, and VS-Code plugins & 1 \\
Kui & 1 \\
k9s & 2 \\
\bottomrule
\end{tabular}
\end{subtable}
\qquad
\begin{subtable}[t]{0.3\linewidth}
\caption{"Other tool" for debugging tasks.}
\label{table:otherdebug}
\begin{tabular}{p{0.825\linewidth}c} 
\toprule
Participant response & Votes \\
\midrule
CLI and scripts for testing, setup, and verification & 1 \\
kubectl, Grafana, loki & 1 \\
Kui & 1 \\
Purpose built tools (e.g. troubleshoot.sh) & 1 \\
CLI and web console & 2 \\
k9s & 4 \\
\bottomrule
\end{tabular}
\end{subtable}
\qquad
\begin{subtable}[t]{0.3\linewidth}
\caption{"Other tool" for monitoring tasks.}
\label{table:othermonitor}
\begin{tabular}{p{0.825\linewidth}c} 
\toprule
Participant response & Votes \\
\midrule
CLI and web console & 1 \\
CLI and scripts for testing, setup, verification, and data collection for analysis and comparisons & 1 \\
Custom application monitoring & 1 \\
kubectl, Grafana, loki & 1 \\
Kui & 1 \\
k9s & 3\\
\bottomrule
\end{tabular}
\end{subtable}
\label{table:other-combined}
\end{table}

\subsection{Part Two: Comparative User Study}

The four participants of the comparative user study had an average of 20 years of programming experience and spend on average 17 hours a week programming. The participants work in the IBM Hybrid Cloud research group as software engineers. 

\subsubsection{Frequency of and Reasons for Tool Modality Use}

We asked participants how often they use a CLI when programming. Two said often and two said always. When asked to list the CLI they most prefer, the most frequently mentioned were kubectl (three of the four) and git (two of the four). Participants shared reasons for why they like or don't like using a CLI, as listed in~\autoref{table:preferencesCLI}. We asked study participants how often they use an IDE when programming. Three said that they always use an IDE and one said that they sometimes use an IDE. All of the participants said Visual Studio Code (VS Code) was their preferred IDE, with two additionally stating Eclipse and IntelliJ IDE.

Participants shared reasons for why they like or don't like using an IDE, listed in~\autoref{table:preferencesIDE}. Participants were asked how often they use a web console when programming. Three said that they rarely use a web console and one said that they never use a web console. Participants were asked to specify which web console they prefer to use and one respondent indicated the OpenShift console. Participants shared reasons for why they like or don't like using a web console, listed in~\autoref{table:preferencesWebCon}. 

\subsubsection{Concurrent Tasks}

Participants shared how they organize their work space when completing concurrent tasks using a web console or CLI, choosing all that apply from the following options: one window, multiple tabs, multiple windows. When performing concurrent tasks using a web console, participants indicated that they use either one window (mentioned by one), multiple windows (mentioned by one), and/or multiple tabs (mentioned by three). One participant indicated that they do not perform concurrent tasks using a web console. When performing concurrent tasks using a CLI, participants indicated that they use either one window (mentioned by two), multiple windows (mentioned by two), and/or multiple tabs (mentioned by four). More windows and tabs are employed when using a CLI to perform concurrent tasks than when using a web console.

\subsubsection{Likert-Scale Evaluation}

All four user study participants ranked the kubectl CLI as the tool interface they preferred to use to complete the task. Participants shared their recommendations for how either interface could be improved See~\autoref{table:recommendations}. Three of the four participants also gave NASA-TLX ratings on a seven-point Likert-Scale from 1 (very low) to 7 (very high) for the kubectl CLI and the OpenShift console to indicate how successful they felt in accomplishing what they were asked to do, how hard they had to work (mentally and physically) to accomplish their level of performance, and how insecure, discouraged, irritated, stressed, and annoyed they were. 

\begin{table}
    \caption{Four user study participants (P1--P4) were asked to reflect on their reasons for preferring a tool modality.}
    
    \begin{subtable}[t]{0.3\linewidth}
        \centering
        \caption{Reasons for CLI preference.}
        \label{table:preferencesCLI}
        \begin{tabular}{lp{0.875\linewidth}} 
            \toprule
             & Participant responses \\ 
            \midrule
            P1 & Succinct, fast \\ 
            P2 & Easier \\
            P3 & Can accelerate tasks but sometimes has limitations and require lot of customization and configuring \\
            P4 & It's hard to remember all possible command/flag combination. Overwhelming number of commands \\
            \bottomrule
        \end{tabular}
    \end{subtable}
\qquad
    \begin{subtable}[t]{0.3\linewidth}
        \centering
        \caption{Reasons for IDE preference.}
        \label{table:preferencesIDE}
        \begin{tabular}{lp{0.875\linewidth}}
            \toprule
             & Participant responses \\ 
            \midrule
            P1 & Code highlights, integrated project management features, debugger \\ 
            P2 & Fewer number of steps to write and test code \\
            P3 & Can accelerate tasks but sometimes has limitations and require lot of customization and configuring  \\
            P4 & Focus on coding and do less grinding work like code navigation, compilation, etc. \\
            \bottomrule
        \end{tabular}
    \end{subtable}
\qquad
    \begin{subtable}[t]{0.3\linewidth}
        \centering
        \caption{Reasons for web console preference.}
        \label{table:preferencesWebCon}
        \begin{tabular}{lp{0.875\linewidth}} 
            \toprule
             & Participant responses \\ 
            \midrule
            P1 & Sometimes it is quicker to use CLI \\ 
            P2 & Prefer the terminal for more control \\
            P3 & Too limiting + leaky abstractions \\
            P4 & Not on par with local IDE in term of features and performance \\
            \bottomrule
        \end{tabular}
    \end{subtable}
\end{table}

\begin{table}
\caption{The recommendations of four user study participants (P1--P4) for improving the kubectl CLI and OpenShift Console. Participant P2 chose not to answer to these questions in the follow-up interview.}
\label{table:recommendations}
\begin{tabular}{lp{0.475\linewidth}p{0.475\linewidth}} 
\toprule
 & kubectl CLI & OpenShift Console \\ 
\midrule
P1 & Cryptic syntax for many commands, hard to do easy things like switch contexts & More developer focused tools \\ 
P2 & -- &  -- \\
P3 & Seems to be doing what it is supposed to do. Minor problems can be super annoying though like this one: \url{https://github.com/kubernetes/kubernetes/issues/42552} & More scripting and automation? Plugins? \\
P4 & Better documentation & NA (not using it) \\
\bottomrule
\end{tabular}
\end{table}

Regarding their success in accomplishing the task using the kubectl CLI, all three gave a rating of 2, and regarding their success using the OpenShift console, one gave a rating of 1, another a rating of 2, and one gave a rating of 4. Regarding how hard they had to work using the kubectl CLI and the OpenShift console, participants gave ratings of 2, 4, and 5, and 3, 4, and 6, respectively. Regarding how insecure, discouraged, irritated, stressed, and annoyed they were completing the task using the CLI and console, all participates gave ratings of 2 for the CLI while two participants rated the console a 3 and one participant rated the console a 4. Therefore, participants found the CLI to have less of an impact on the effort required to perform the task and overall felt more successful using the CLI.

\section{Discussion and Design Implications}

This paper investigates user {\em preference} and {\em productivity} in Cloud tooling, with an eye towards finding gaps and opportunities in the way these tools are designed. The level of design commitment seems to vary widely across these modalities. Consoles are heavy with design intent, rich with functionality across the provider's Cloud offerings, and are often backed by large design and development teams. CLIs are typically quite low level and supported by small teams. However, is this the best balance? From a design perspective, there are a variety of consoles being maintained and developed to benefit “low/no-code” \cite{mclean2021, richardson2016}. It is uncertain whether this high investment is returning a set of tools that developers prefer to use. In order to more deeply understand the impact that these tool modalities have on productivity when interacting with the Cloud, we must consult the developers that use these tools in their everyday work. The number of other tools that were mentioned among survey respondent preferences indicates the prevalent diversity of Cloud tooling. See~\autoref{table:other-combined}.

A vital part of the developer’s user experience today is how knowledge can be captured, shared, and more effectively communicated. Research has shown that using CLIs is the most effective way for developers to learn and gain experience. The external resources that exist to teach people how to use a CLI generally don’t make as much use of screenshots in their tutorials and therefore don’t require as much maintenance effort as documentation for other tool modalities such as IDEs \cite{kubectl2022, gitdocs2022, openshift2022, intellijdocs2022, vscode2022}. A steeper learning curve remains associated with CLIs, despite their consistent and necessary role in development. During the observational user study, when participants completed the task using the CLI they would often repeatedly execute commands due to spelling errors. Only two of the participants made use of keyboard shortcuts when using the CLI. Furthermore, study participants would remark that they could not remember specific command combinations to efficiently execute several steps of the task at once. 

Changes to design focus are needed to impart the level of productivity that daily use of tooling requires. Initially, we thought beginners would prefer graphical over textual tool modalities and while three beginner survey respondent preferences deviated from CLIs, none indicated they preferred an IDE over a CLI or web console. Only three survey respondents indicated they preferred an IDE, all intermediate developers. Based on the findings presented in this paper that show developers prefer CLIs to accomplish CRUD and debugging, and web consoles to perform monitoring, and given the daily frequency that developers accomplish multiple types of tasks, we recommend exploring the impact of more integrated, hybrid tool modalities to avoid the cost of context switching. We have conducted some preliminary research for a follow-up to this study to further explore comparable measures of productivity across these disparate tool modalities. By quantifying the interactions required to use each tool, we seek to be able to compare the manual effort required to complete a task.

\section{Threats to Validity}
In light of the ongoing push towards automation, a challenge to this study is against our focus on tasks that users perform manually. This is despite an ongoing push for developers to employ automation in favor of manual tasks; e.g. GitOps~\cite{gitops} and CI/CD~\cite{cicd} more generally. This is a reasonable critique, however, as~\autoref{fig:task-frequency} shows, people are still heavily reliant on manual tools to accomplish their routine tasks.

A challenge we faced was in deciding what metrics to use to compare these tools in our user study. We assume that these tools are already well designed and instead seek to study the limits that exist within the steady state of design and learning. These modalities will have certain limits and we seek to understand what are they as is in order to determine how we can improve them. We found that wall clock time is not the best comparison because of the user variance we observed when participants completed a task in preliminary pilot testing. User preference is a better metric, however, it may be too subjective. For example, knowing people prefer a CLI for certain tasks does not help to improve the web console for those tasks.

One person from the open source community gave us feedback on our survey questions and said that OpenShift is a distribution and not a Cloud provider. Perhaps not including OpenShift as an option would have encouraged more specific responses for other Cloud providers not listed as popular options. Furthermore, we only collected five responses from beginners and require more representation from this user group in future work as they are not as equally represented in our participant pool.

\section{Conclusion and Future Work}

CLIs have been with us since the dawn of modern computing and they continue to use comparatively primitive presentation methods. Recall from Section 2 the low-level and tedious CLI interface Star used for her monitoring task. This scenario illuminates the opportunity for a grand re-imagining of the Cloud user experience, employing modern and sleek graphical toolkits. Large companies such as RedHat have clearly thrown considerable resources into this effort, in developing their OpenShift console. However, we have found that most developers only prefer to use the web console for monitoring tasks. In general, most developers prefer to use the CLI to accomplish CRUD and debugging tasks. 

We as a community have yet to find something truly better than a CLI. Hybrid interfaces are a promising direction for future work in this area. We hope to explore the impact of hybrid interfaces on developer tool preference in the context of different types of programming tasks. We began preliminary pilot testing of a user study with two participants that used Kui, a hybrid tool to accomplish the same task that our user study participants carried out using the kubectl CLI and the OpenShift console. Furthermore, we hope to study means for improving the quantification of user preference with the goal of constructively informing the design of Cloud tools. We identified a set of comparable metrics from screen recordings of the comparative user study task sessions to better quantify participant exhaustion and disruption and reinforce self-reported NASA-TLX scale ratings.

\section{Acknowledgments}
This research was conducted as part of a summer internship with the IBM T.J. Watson Research Center.

\bibliographystyle{ACM-Reference-Format}
\bibliography{sample-base}

\end{document}